\documentclass[12pt,preprint]{aastex}
\usepackage{natbib}
\usepackage{graphicx}
\def\apjs{{\rm ApJS}}                  
\def\apjl{{\rm ApJ Let}}               

\def\Msol{\thinspace\hbox{$\hbox{M}_{\odot}$}}

\def\a4{\hsize 17.0cm \vsize 25.cm}
\defcitealias{Draine2011}{Dr11} 
\defcitealias{SilichTenorioTagle2013}{ST13} 
\defcitealias{Weaveretal1977}{Weaver's et al.}

\def\Msol{\thinspace\hbox{$\hbox{M}_{\odot}$}}

\shorttitle{Ionized Shells}
\shortauthors{Mart\'{\i}nez-Gonz\'alez et al.}

\begin{document}

\title{On the impact of radiation pressure on the dynamics and inner structure
       of dusty wind-driven shells.}

\author{Sergio Mart\'{\i}nez-Gonz\'alez  \altaffilmark{1},
Sergiy Silich \altaffilmark{1},
Guillermo  Tenorio-Tagle\altaffilmark{1}
}

\altaffiltext{1}{Instituto Nacional de Astrof\'\i sica \'Optica y
Electr\'onica, AP 51, 72000 Puebla, M\'exico}

\begin{abstract}
Massive young stellar clusters are strong sources of radiation and mechanical 
energy. Their powerful winds and radiation pressure sweep-up interstellar gas 
into thin expanding shells which trap the ionizing radiation produced by the 
central clusters affecting the dynamics and the distribution of their ionized 
gas. Here we continue our comparison of the star cluster winds and radiation 
pressure effects on the dynamics of shells around young massive clusters. We 
calculate the impact that radiation pressure has on the distribution of matter
and thermal pressure within such shells as well as on the density weighted 
ionization parameter $U_w$ and put our results on the diagnostic diagram which
allows one to discriminate between the wind-dominated and radiation-dominated 
regimes. We found that model predicted values of the ionization parameter 
agree well with typical values found in local starburst galaxies. Radiation 
pressure may affect the inner structure and the dynamics of wind-driven shells
significantly but only during the earliest stages of evolution (before 
$\sim 3$~Myr) or if a major fraction of the star cluster mechanical luminosity
is dissipated or radiated away within the star cluster volume and thus the 
star cluster mechanical energy output is significantly smaller than star 
cluster synthetic models predict. However, even in these cases radiation 
dominates over the wind dynamical pressure only if the exciting cluster is 
embedded into a high density ambient medium.

\end{abstract}


\section{Introduction}
\label{sec:1}

HII regions are fundamental to our understanding of young stellar clusters 
radiative and mechanical feedback on the interstellar medium (ISM). They are 
strong sources of emission-line radiation and thus serve as a powerful 
diagnostic tool to study star formation and the chemical composition of 
nearby and distant galaxies \citep{CapriottiKozminski2001,Dopitaetal2005,
Dopitaetal2006, YehMatzner2012}. They have even been used as tracers of the 
Hubble expansion \citep{Chavezetal2012}. The idealized \citep{Stromgren1939} 
model for spherical static HII regions with a homogeneous density distribution
was a revolutionary step forward in the study of photoionized nebulae. However the
consideration of a number of physical effects have led to a much more robust 
paradigm. Winds produced  by the exciting clusters 
\citep{CapriottiKozminski2001,Arthur2012,SilichTenorioTagle2013} and the 
impact that radiation pressure provides on the swept-up interstellar gas 
\citep{ElmegreenChiang1982, CapriottiKozminski2001, Matzner2002, 
KrumholzMatzner2009, NathSilk2009, SharmaNath2012} are among such 
major physical effects. As recently shown by \citet{Draine2011}, the 
absorption of photons emerging from an exciting cluster by either dust grains 
and recombining atoms, leads to a non homogeneous density distribution even 
within static or pressure confined  HII regions and under certain conditions, radiation pressure 
may pile up the ionized gas into a thin outer shell, as assumed by 
\citet{KrumholzMatzner2009}. The action of cluster winds, as well as the strong evolution
that the ionizing photon flux and the star cluster bolometric luminosity 
suffer after the first supernova explosion make the situation even more 
intricate \citep{SilichTenorioTagle2013}.

The thermalization of the stellar winds and supernovae mechanical energy 
through nearby random collisions leads to a high central overpressure which
forms a strong shock that moves supersonically and sweeps the ambient 
ionized gas into a thin, wind-driven shell. This shell cools down in a short 
time scale and begins to absorb ionizing photons causing the ionization front 
to move back towards the cluster and finally become trapped within the shell. 
The size and density distribution of such ionized shells have little to do 
with the original Str\"omgren model. Their evolution depends not only on the 
ambient gas density distribution and the available Lyman continuum, but also 
on the mechanical power of the exciting cluster. \citet[hereafter ST13]{SilichTenorioTagle2013}
discussed the impact that radiation pressure has on the dynamics of  
wind-driven shells powered by young star clusters and found radiation pressure 
not to be a dominant factor. They,
however, did not consider the detailed impact that radiation pressure provides
on the inner shell structure. They also assumed that shells absorb all photons
escaping from the central cluster and thus found an upper limit to the 
radiative feedback from the central cluster on the dynamics of the swept-up 
shell. Here we extend the analysis provided in 
\citetalias{SilichTenorioTagle2013} and discuss how radiation pressure affects
the distribution of density and thermal pressure within a shell and thus how 
it may affect the velocity of the outer shock and the dynamics of the ionized 
gas around young stellar clusters. 

The paper is organized as follows: we first present in section \ref{sec:2} 
the major equations formulated by \citet{Draine2011} for static spherically 
symmetric HII regions and discuss how the inner and outer boundary conditions 
affect the solution. In section \ref{sec:3} we discuss different hydrodynamic 
regimes and also show how \citetalias{Draine2011}'s equations may be applied to the whole 
shell, including the outer, non-ionized segments. The results of the 
calculations are presented and discussed in section \ref{sec:4} where we 
compare different hydrodynamical models (standard energy and momentum 
dominated, leaky and low star clusters heating efficiency), calculate the 
model-predicted values of the ionization parameter and compare them to typical 
values found in local starburst galaxies. Our results are also placed onto a 
diagnostic diagram which allows one to discriminate between the radiation 
pressure and wind pressure (thermal or ram) dominated regimes. The summary of 
our major results is given in section \ref{sec:5}.        

\section{Radiation pressure in static, dusty HII regions}
\label{sec:2}

Let us first consider the idealized model of a static spherically symmetric 
HII region ionized by a central star cluster and confined by the thermal 
pressure of the ambient interstellar medium (ISM). Following \citet[hereafter Dr11]{Draine2011}, 
we assume that the outward force provided by radiation pressure is balanced by the inward directed 
thermal pressure gradient. The set of equations describing such HII regions in the presence of 
dust grains is \citepalias[see][]{Draine2011}:
\begin{eqnarray}
\label{eq:equilibrium}      \hspace{-0.4cm}
\frac{d}{dr}\left(\frac{\mu_i}{\mu_a} n k T_i\right) &=& n\sigma_d \frac{\left[L_n e^{-\tau}+L_i \phi\right]}{4\pi r^2 c} + 
n^2 \beta_2 \frac{\langle h\nu\rangle_i}{c} ~,\\
          \label{eq:phi}      \hspace{-0.4cm}
S_{0}\frac{d\phi}{dr} &=&  - \beta_2 n^2  - n\sigma_d S_{0}\phi  ~,\\     
          \label{eq:tau}      \hspace{-0.4cm}
\frac{d\tau}{dr} &=& n\sigma_d ~,
\end{eqnarray}
where $L_i$ and $L_n$ are the luminosities in ionizing and non-ionizing photons, 
respectively ($L_i+L_n=L_{bol}$, where $L_{bol}$ is the bolometric luminosity of
the cluster), $n(r)$ is the ionized gas density, $\phi(r)$ is the fraction of the 
ionizing photons that reaches a surface with radius $r$, $S_{0}=Q_{0}/4\pi r^2$ 
where $Q_0$ is the number of ionizing photons emitted by the star cluster per second,
$\langle h\nu\rangle_i = L_i/Q_0$ is the mean energy of the ionizing photons, 
$\tau(r)$ is the dust absorption optical depth, $\sigma_d$ is the effective dust 
absorption cross section per hydrogen atom, $\beta_2 = 2.59 \times 10^{-13}$~cm$^3$ 
s$^{-1}$ is the recombination coefficient to all but the ground level \citep{Osterbrock1989}, 
$k$ and $c$ are the Boltzmann constant and the speed of light, respectively, and 
$T_i$ is the ionized gas temperature. It is assumed that the gas in the HII region 
is completely ionized and has a normal chemical composition with one helium atom per 
every ten hydrogen atoms. The mean mass per particle and the mean mass per ion then are: 
$\mu_a = 14/23 m_{H}$ and $\mu_i = 14/11 m_{H}$, respectively, where $m_H$ is the proton mass. 
We set the value of the dust absorption cross section per hydrogen atom to $\sigma_d=10^{-21}$ cm$^{2}$ 
\citepalias{Draine2011} and assumed that the temperature of the ionized gas is constant 
and equal to $T_{i}= 10^4 \mbox{ K}$ in all our calculations. The first and the second 
terms on the right-hand side of equation (\ref{eq:equilibrium}) correspond to the 
photon momentum absorbed by dust grains and by the gas, respectively. 
The right-hand terms in equation \ref{eq:phi} are the rates of absorption of ionizing 
photons in a thin spherical shell with radius $r$ and thickness $dr$ by recombination and by 
dust grains, respectively.

In order to select a unique solution of equations (\ref{eq:equilibrium} - 
\ref{eq:tau}), one has to adopt a set of initial or boundary conditions. 
For example, \citet{Draine2011} selected solutions by choosing the initial 
value of density at some fixed radius $r$. We use similar initial conditions
in the case of the wind-driven shell (see next section), but prefer to select
the static solution from the two boundary conditions which are the values of 
the confining pressure at the inner and outer edges of the HII region. 
Here we assume that the HII region is static and that the radiation field from the 
central cluster is strong enough to clean up the central region with a radius $R_i$
as it seems appropriate to many galactic and extragalactic HII regions which are better
fitted with models containing an empty central zone in the ionized gas distribution 
\citep[see][]{Mathews1967,Mathews1969,KewleyDopita2002,Dopitaetal2003}. In such a case, 
the conditions at the inner edge of the HII region are: $\phi(R_i)=1$, $\tau(R_i)=0$ 
and $n(R_i) \rightarrow 0$ \citepalias[see][]{Draine2011} whereas the value of the 
initial radius $R_{i}$ is selected by the outer boundary condition which requires the 
thermal pressure at the outer edge of the HII region $R_{HII}$ to be equal to that 
in the ambient ISM.
We use in the calculations a value of $n(R_i)=10^{-10}$ cm$^{-3}$ and stop the
integration when all ionizing photons are trapped and thus the function $\phi$
becomes equal to zero: $\phi(R_{HII})=0$. 

The input parameters ($Q_0$, $L_i$ and $L_n$) for our calculations were taken from 
the Starburst99 synthesis model \citep{Leithereretal1999} and are summarized in 
Table \ref{tab:1}. Models A, B and C correspond to a $10^6$ M$_\sun$ coeval 
stellar cluster with a standard Kroupa initial mass function with 
upper and lower cut-off mass of 100\Msol \, and 0.1\Msol\, respectively and a 
turn off mass at 0.5\Msol, metallicity $Z= 0.4$ Z$_\sun$, age $t \sim 
1$~Myr and Padova evolutionary tracks with AGB stars, embedded 
into an interstellar gas with number density 1~cm$^{-3}$, $10^3$~cm$^{-3}$ and
$10^6$~cm$^{-3}$, respectively. Models D, E and F correspond to a two orders 
of magnitude less massive cluster of the same age located within the same 
environments.
\begin{table}[htp]
\caption{\label{tab:1} Stationary HII region models}
\begin{tabular}{c c c c c}
\hline\hline
Models  & $Q_0$ & $L_i$ & $L_n$ & $n_{ISM}$ \\
        & s$^{-1}$ & erg s$^{-1}$ &  erg s$^{-1}$ & cm$^{-3}$ \\\hline
A,B,C   &  $4.27\times 10^{52}$ & $1.44\times 10^{42}$   & $1.96\times 
10^{42}$ & 1, 10$^3$, 10$^6$ \\
D,E,F   &  $4.27\times 10^{50}$ & $1.44\times 10^{40}$   & $1.96\times 
10^{40}$ & 1, 10$^3$, 10$^6$ \\
\hline\hline
\end{tabular}
\end{table}
\begin{figure*}
\plotone{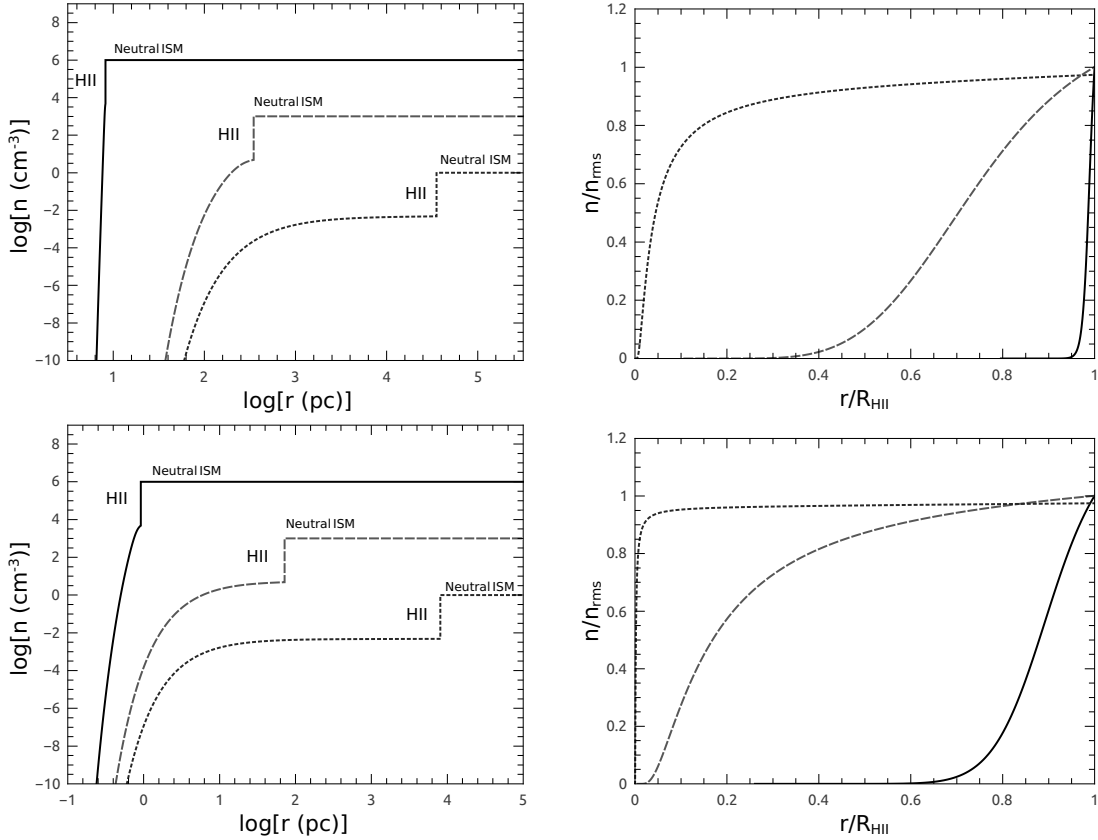}
\caption{Static HII regions with a central cavity. The upper left-hand panel 
presents the gas number density distribution as a function of radius for 
models A (dotted line), B (dashed line) and C (solid line) in a log-log 
scale. The upper right-hand panel shows the same distributions when all 
distances are normalized to the radius of the HII region and densities to 
their rms values. The bottom panels present similar density distributions 
for  models D (dotted line), E (dashed line) and F (solid line), 
respectively.}
	\label{fig:1}
\end{figure*}

The calculated density distributions for static HII regions with a central 
cavity are shown in Figure \ref{fig:1}. The density  grows always rapidly in a
very narrow inner zone and then presents an almost even or flat distribution 
in the rest of the volume if the density of the ambient ISM is not very large 
(models A, B, D and E). Only when the exciting clusters are embedded into a 
very high density ambient medium ($n_{ISM}=10^6$ cm$^{-3}$, models C and F) 
the density of the ionized gas grows continuously across the whole HII region. 
However, such HII regions are very compact (see left-hand panels in Figure 
\ref{fig:1}). The size of the HII region, $R_{HII}$, and the radius of the 
inner empty cavity $R_i$ are both functions of the interstellar ambient 
density. Both radii grow rapidly as one considers a lower ambient density (see
the left-hand side panels in Figure \ref{fig:1} where the steps in the gas 
density distribution mark the edge of the HII regions and result from the 
condition that the thermal pressure at the HII region edge ought be equal 
to that of the ambient neutral gas with a two orders of magnitude lower 
temperature ($T_{ISM}=100$ K). This however is not evident when distances are 
normalized to the radius of the HII region and densities to their rms values
as on Figure 2 of \citetalias{Draine2011} (see right-hand panels in
Figure \ref{fig:1}). Thus dimensionless plots do not allow one to
realize that static models with a low ambient density are unrealistic  
as in these cases the required time for the ionized gas 
re-distribution (the sound crossing time) highly exceeds the
characteristics life time of the HII region, $\tau_{HII} \sim 10$~Myr.

\section{Radiation pressure in dusty wind-driven shells}
\label{sec:3}

Given the continuous supply of photons and their instantaneous re-processing 
by the surrounding gas here we use \citetalias{Draine2011}'s equations 
to calculate the impact that radiation pressure has on the structure and on 
the dynamics of evolving wind-driven shells. We consider a constant density 
ISM and a set of evolving star cluster parameters to evaluate at consecutive 
times the impact of radiative pressure on the evolving shells and thus neglect
all effects dealing with a plane-stratified density distribution in galactic 
disks, gas shear and gravity which were thoroughly discussed in our previous
papers \citep[see, for example,][]{TenorioTaglePalous1987,Silich1992,Silichetal1996} 
and do not present the major aim of this paper.
We also do not consider the impact that the ambient pressure 
provides on the shell dynamics as it is only significant when the shell 
expansion velocity approaches the sound speed value in the ambient ISM. The 
distribution of the ionized gas then becomes quasi-static and is defined by 
the values of thermal pressure at the inner edge of the shell and in the 
ambient ISM as was discussed in the previous section. In the supersonic 
regime, which is the case in all our calculations (see Mach number 
values in the captions to Figures \ref{fig:3} and \ref{fig:4}, calculated
under the assumption that the sound speeds in the ionized and neutral 
ISM are 15~km s$^{-1}$ and 1.04~km s$^{-1}$, respectively), the rate of mass 
accumulation by the expanding shell depends on the speed of the leading shock,
$V_s \sim (P_{edge}/\rho_{ISM})^{1/2}$, where $P_{edge}$ is the thermal 
pressure value immediately behind the leading shock
and $\rho_{ISM}$ is the gas density in the ambient ISM. The impact of the
external pressure on the shell dynamics is thus negligible in this case.
When a star cluster wind impacts a constant density ISM, a four zone structure 
is established: there is a central free wind zone, surrounded by a shocked 
wind region. The latter is separated by a contact discontinuity from the 
matter swept up by the leading shock  which evolves into the constant density 
ISM \citep[see][]{Weaveretal1977, MacLowMcCray1988, KooMcKee1992}. 
In the wind-blown bubble case, the central zones are hot and thus transparent 
to the ionizing flux as it is also the case in the static HII regions with a 
central cavity considered in the previous section. However, the density at 
the inner edge of the ionized shell is not arbitrarily small, but must be 
selected from the condition that $P_{HII}(R_{s}) = P_s$, where 
$P_{HII}(R_{s})$ and $P_s$ are the thermal pressures at the inner edge of the 
ionized shell and in the shocked wind region, respectively, and $R_{s}$ is 
the radius of the contact discontinuity (the inner radius of the ionized 
shell). 
The swept up shell is also hot at first ($T \ge 10^6$ K) and thus 
transparent to the ionizing radiation from the star cluster. However it  cools
down in a short time scale due to strong radiative cooling. If the density 
and metallicity of the ambient medium are $n_{ISM}$ and $Z_{ISM}$, 
respectively, and the star cluster mechanical luminosity is $L_{mech}$, the 
shell characteristic cooling time scale, $\tau_{cool}$, is 
\citep{MacLowMcCray1988}:
\begin{equation}
      \label{eq6}
\tau_{cool} = (2.3 \times 10^4) Z^{-0.42}_{ISM} n^{-0.71}_{ISM} 
              \left( \frac{L_{mech}}{10^{38} 
\mbox{ erg} \mbox{ s}^{-1}}\right)^{0.29} yr .
\end{equation}
Only after that time the swept up shell begins to recombine and absorb the 
ionizing radiation from the central cluster. For $L_{mech}=10^{40}$ erg 
s$^{-1}$, $Z_{ISM}=0.4 Z_{\sun}$ and $n_{ISM}=1$ cm$^{-3}$, $\tau_{cool} 
\sim 0.12$~Myr while for an ISM with $n_{ISM}=1000$ cm$^{-3}$, $\tau_{cool} 
\sim 10^{-3}$~Myr. 
When the wind-driven shell grows thick enough, it absorbs all ionizing photons
and then forms an outer neutral skin which absorbs only non-ionizing photons 
that manage to escape the inner ionized part of the shell. One can calculate 
how these photons affect the distribution of density and thermal pressure in 
the neutral part of the shell by removing the ionizing radiation from 
\citetalias{Draine2011}'s equations and evaluating the rate at which non-ionizing energy, 
$L_{nesc}$, escapes from the ionized part of the shell. This leads to the 
set of equations: 
\begin{eqnarray}
          \label{eq:equi_neu}      \hspace{-0.4cm}
\frac{d n}{dr}  &=& \frac{n\sigma_d}{k T_n} 
\frac{L_{nesc} e^{-\tau}}{4\pi r^2 c} ~,\\     
          \label{eq:tau_neu}      \hspace{-0.4cm}
\frac{d\tau}{dr} &=& n\sigma_d ~.
\end{eqnarray}
It was assumed in all calculations that the temperature in the outer, neutral 
part of the shell is constant and equal to $T_n = 100$~K. It was also assumed 
that the shell is thin and thus the total mass of the shell is $M_{sh} = 4 \pi 
\rho_{ISM} R_{s}^3 / 3$. 

Thus, the initial conditions which allow one to select a unique solution of 
equations (\ref{eq:equilibrium}-\ref{eq:tau}) in the case of the wind-blown 
shell are very similar to those used in the previous section: 
$\phi(R_{s}) = 1$, $\tau(R_{s})=0$ and the value of the thermal 
pressure in the shocked wind zone which depends upon the dynamical time $t$.
However the inner radius of the ionized shell $R_{s}$ and the pressure 
$P_s$ at the inner edge of the shell at different evolutionary times $t$ 
are calculated from the \citet{Weaveretal1977} wind-blown bubble model and 
the integration stops when the total mass of the ionized and neutral segments
reaches $M_{sh}$. We then compare the values of thermal pressures at the outer 
($P_{edge}$) and inner ($P_s$) edges of the swept-up shell obtained from the
calculations in order to check if radiation pressure may affect the shell 
dynamics significantly.

In the energy-dominated regime, $R_{s}$ and $P_{s}$ are \citep{BKSilich1995}
\begin{eqnarray}
      \label{eq2a}
      & & \hspace{-1.1cm} 
R_{s}(t) = \left[\frac{375(\gamma-1) L_{mech}}
                      {28(9\gamma-4)\pi\rho_{ISM}}\right]^{1/5} t^{3/5} ,
      \\[0.2cm]     \label{eq2b}
      & & \hspace{-1.1cm}
P_{s}(t) = 7 \rho^{1/3}_{ISM} \left[\frac{3(\gamma-1) L_{mech}}
                  {28(9\gamma-4) \pi R^2_{s}}\right]^{2/3} ,
\end{eqnarray}		
where $\rho_{ISM}$ is the interstellar gas density and $\gamma=5/3$ is the ratio of specific heats.
At this stage the free wind occupies only a small fraction of the bubble volume and the value of
thermal pressure $P_{s}$ is defined by the amount of thermal energy accumulated
in the shocked wind region and the bubble volume and thus does not depend on
the wind terminal speed \citep[see for more details][]{BKSilich1995}.

The ion number density at the inner edge of the ionized shell then is:
\begin{equation}
      \label{eq2c}
n_{s}(t) = \frac{\mu_a P_{s}}{\mu_i k T_i} .
\end{equation}

However, evaporation of the swept-up shell into the hot shocked wind region 
may cause strong radiative cooling and lead to the end of the energy dominated
regime. If the star cluster is embedded into an ambient ISM with density 
$n_{ISM}$, this occurs at \citep{MacLowMcCray1988}:
\begin{equation}
      \label{eq15}
\tau_{tran} = (1.6 \times 10^7) Z^{-35/22}_{ISM} n^{-8/11}_{ISM} 
              \left( \frac{L_{mech}}{10^{38} \mbox{ erg} \mbox{ s}^{-1}}\right)^{3/11} yr .
\end{equation}
After this time the free wind impacts directly on the shell and the thermal 
pressure at the inner edge of the swept-up shell is equal to the wind ram 
pressure $P_{ram}=\rho_w V_{\infty}^2$, where $\rho_{w} = \dot{M}_{SC}/4\pi 
R_s^2 V_{\infty}$, $\dot{M}_{SC}$ is the star cluster mass deposition rate and 
$V_\infty=(2L_{mech}/\dot{M}_{SC})^{1/2}$ is the adiabatic wind terminal 
speed. The shell further expands in the momentum dominated regime as 
\citepalias[see][]{SilichTenorioTagle2013}:
\begin{eqnarray}
      \label{eq13a}
      & & \hspace{-1.1cm} 
R_{s}(t) = R_{tran}
\left[\frac{3 L_{mech} (t^2 + \tau_{tran}^2)}
      {\pi V_{\infty} \rho_{ISM} R^4_{tran}} +
      \left(\frac{12}{5} - \frac{6 L_{mech} \tau_{tran}^2}
      {\pi V_{\infty} \rho_{ISM} R^4_{tran}}\right) \frac{t}{\tau_{tran}} -
      \frac{7}{5}\right]^{1/4} \, ,
      \\[0.2cm]     \label{eq13b}
      & & \hspace{-1.1cm}
P_{s}(t) = \frac{L_{mech}}{2 \pi V_{\infty} R^2_{s}} ,
\end{eqnarray}
The radius of the shell at the time of the transition, $R_{tran}$, must be 
calculated by means of equation (\ref{eq2a}) at $t=\tau_{tran}$.
The above equations do not include the momentum of starlight. However, 
as shown below, this does not make a major difference in the evolution of the 
wind-blown shells except in the case of a low heating efficiency. Our 
calculations thus allows one to realize when the standard model assumptions may
break down and radiation pressure may affect the dynamics of the wind-driven 
shells. The ion number density at the inner edge of the ionized shell in 
this case is:
\begin{equation}
      \label{eq13c}
n_{s}(t) = \frac{\mu_a P_{ram}}{\mu_i k T_i} .
\end{equation}
Note, however, that if thermal conduction and mass evaporation of the 
outer shell are inhibited by magnetic fields, the radiative losses of energy 
from the shocked wind region remain negligible and the wind-driven bubble 
expands in the energy dominated regime during the whole evolution of the HII 
region \citep[see][]{SilichTenorioTagle2013}.

\begin{figure*}[htp]
\plotone{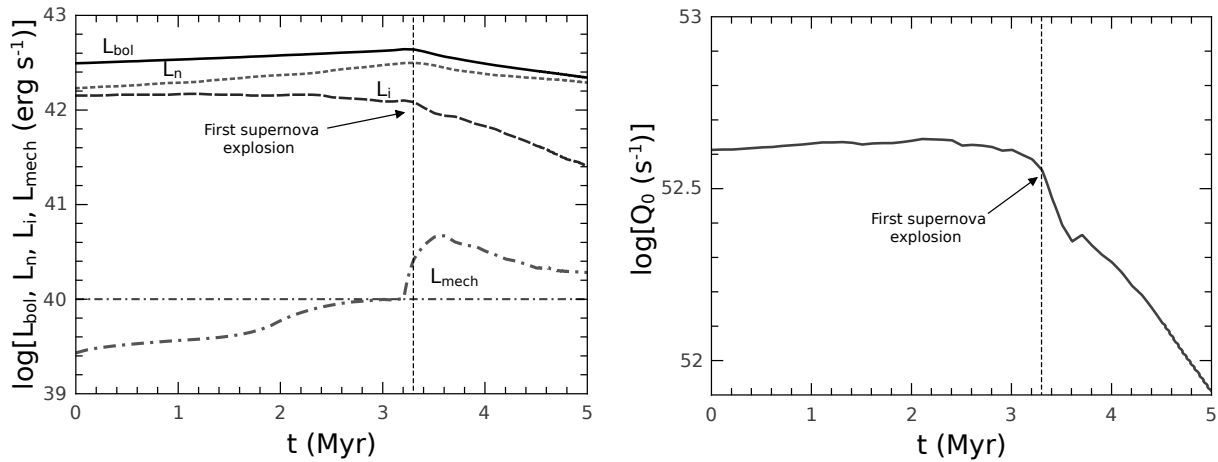}
\caption{Input parameters as a function of time. The left-hand panel shows 
the evolution of the bolometric (solid line), non-ionizing (dotted line),
ionizing (dashed line) and mechanical (dash-dotted line) luminosities. 
The horizontal dash-dotted line displays the value of mechanical luminosity
that has been used in \citet{Weaveretal1977} analytic relations. The 
right-hand panel shows the number of ionizing photons produced by a 
$10^6$\Msol \, cluster per unit time. The vertical lines in both panels mark 
the onset of supernova explosions.}
\label{fig:2}
\end{figure*}

To evaluate the impact that radiation pressure provides on the shell, we 
run several models (see Table \ref{tab:2}) with input parameters $L_{bol}$, 
$L_{i}$, $L_{n}$, $L_{mech}$ and $Q_{0}$, which one has to know in order to 
use equations (\ref{eq:equilibrium} - \ref{eq:tau}) and (\ref{eq:equi_neu} - 
\ref{eq:tau_neu}), taken from the Starburst99 synthesis code at the 
corresponding times $t = 1$~Myr, 3.3~Myr and 5~Myr, respectively (see Figure 
\ref{fig:2}). 
\begin{table}[htp]
\caption{\label{tab:2} Wind-driven shell models}
\begin{tabular}{c c c c c c c}
\hline\hline
N & Models  & $L_{SC}$ & $n_{ISM}$ & $Z_{ISM}$ & $t$ & Regime       \\
  &      & erg s$^{-1}$ & cm$^{-3}$ & $Z_{\odot}$ & Myr &        \\\hline
1 & LDS a, b, c & $10^{40}$ & 1 & 0.4 & 1, 3.3, 5 & Low density energy dominated  \\
2 & LDL a, b, c & $10^{40}$ & 1 & 0.4 & 1, 3.3, 5 & Low density with gas leakage  \\
3 & HDS a, b, c & $10^{40}$ & $10^3$ & 0.4 & 1, 3.3, 5 & High density energy/momentum dominated \\
4 & HDE a, b, c  &  $10^{40}$ & $10^3$ & 0.4 & 1, 3.3, 5 & High density with low heating efficiency \\
\hline\hline
\end{tabular}
\end{table}
To calculate the inner radius of the ionized shell and the ionized gas density
at the inner edge of the HII region we used equations (\ref{eq2a}-\ref{eq2c}) 
and if equation (\ref{eq15}) is fulfilled these were replaced by equations 
(\ref{eq13a}-\ref{eq13c}). Our procedure thus implies that the dynamical 
evolution of the shell is done through the classical 
\citetalias{Weaveretal1977} equations, while the density and pressure 
structure of the swept up shell is evaluated by means of 
\citetalias{Draine2011}'s static equations. In all of these, an average mechanical luminosity 
$L_{mech} = 10^{40}$ erg s$^{-1}$ was used (see the discussion in section 4.2 
and in \citetalias{SilichTenorioTagle2013}).

\section{Results and discussion}
\label{sec:4}

\subsection{Shells evolving in a low density ISM}

We first explore the impact that radiation provides on the wind-driven shells 
expanding into a low density ambient medium (Table \ref{tab:2}, LDS model). 
Models LDSa, LDSb and LDSc present different evolutionary stages of the 
``standard bubble model''. In this case the wind-driven shell expands into a 
low density 
(1 cm$^{-3}$) ISM in the energy-dominated regime. In all cases the mass of 
the driving cluster is $10^6$\Msol \, and the selected times  allow one to see
how the ionization structure of the shell changes with time due to the bubble 
and radiation field evolution. 

\begin{figure*}[htp]
\plotone{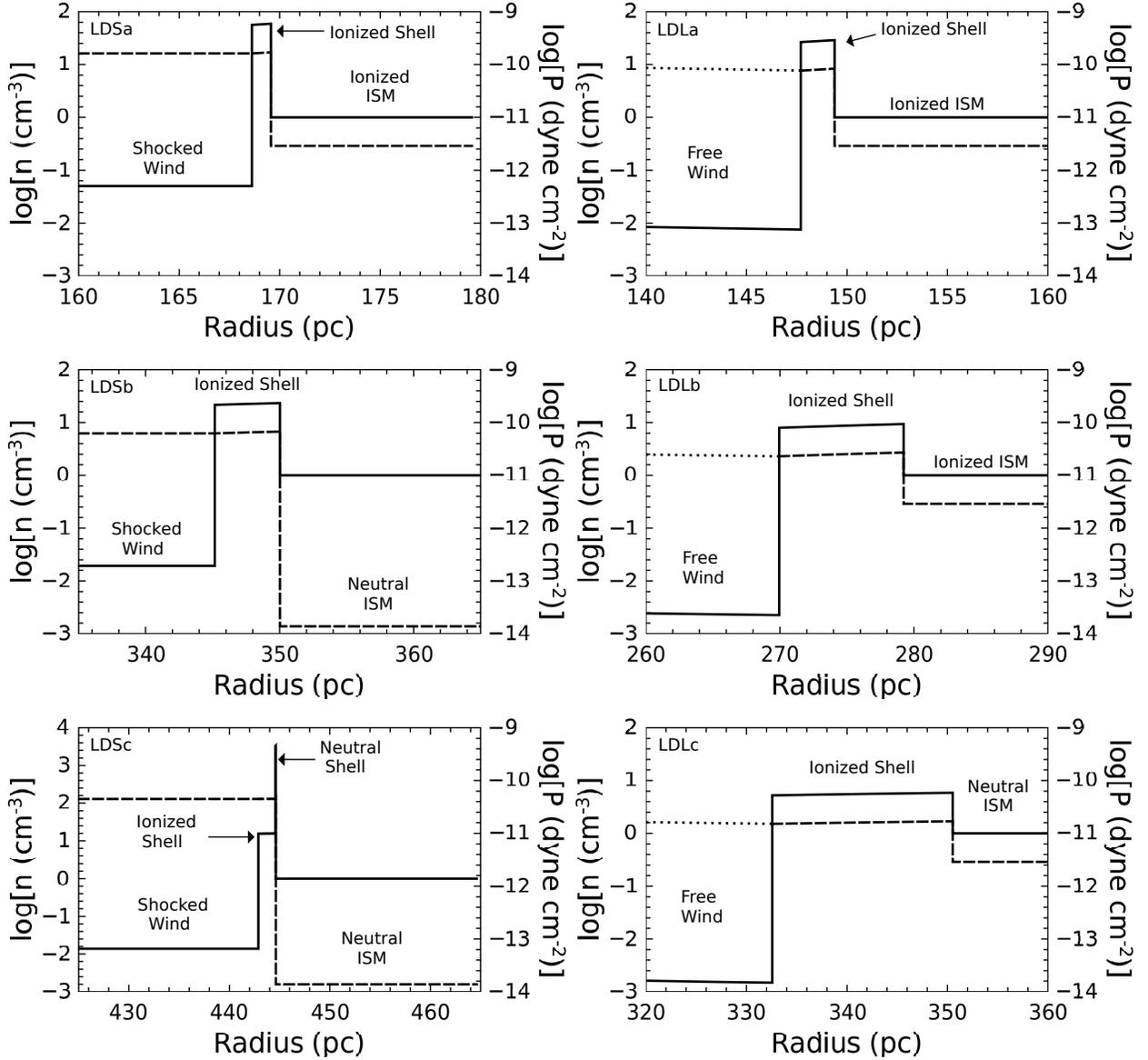}
\caption{The wind-blown shell structure for a low-density environment. Zoom 
at the density (left-hand axes) and pressure (right-hand axes) distributions 
across and at both sides of the expanding shell. The left-hand panels 
present the results of the calculations for models LDSa (top panel), LDSb 
(middle panel) and LDSc (bottom panel). The right-hand panels displays the 
results for models with gas leakage: models LDLa (top panel), LDLb (middle 
panel) and LDLc (bottom panel), respectively. Solid lines show the radial 
density distribution in the shocked/free wind region, in the ionized and 
neutral shell and in the ambient ISM. Dashed and dotted lines display the 
distribution of thermal pressure inside the shell and in the ambient ISM and 
that of the ram pressure in the free wind region, respectively.
The Mach number for the LDS models a, b and c is 6.6, 59 and 50, 
respectively, while for the LDL models a, b and c is 4.9, 2.7 and 31.4.} 
	\label{fig:3}
\end{figure*}

Figure \ref{fig:3} displays the  density (solid lines), thermal pressure 
(dashed lines) and ram pressure (dotted lines) distributions within and at 
both sides of the expanding shell, while this is exposed to the radiation 
from the central cluster.
The sudden density jumps at the inner edge of the ionized shell result from 
the fact that the thermal pressure there must be equal to the thermal pressure
of the hot thermalized cluster wind (equations \ref{eq2c}) while the 
temperature in the ionized gas is $10^4$~K. 
As shown in Figure \ref{fig:3}, in the case of model LDSa ($t = 1$~Myr) the 
swept up shell has  already cooled down and is completely photo-ionized by 
the Lyman continuum from the young central cluster. Furthermore, a  fraction 
of the ionizing photons still escapes from the shell into the ambient ISM 
keeping  it also at $T = 10^4$ K. Model LDSb presents the shell structure at 
the trapping time, $\tau_{trap} = 3.3$~Myr. At this time the shell absorbs 
all ionizing photons, and the mass of the ionized matter is exactly that 
of the swept-up shell: $M_{ion} = M_{sh}$. The thermal pressure outside of 
the shell then falls by two orders of magnitude as it is assumed that the 
temperature of the ambient neutral gas in this case is $100$~K (see the 
left-hand middle and bottom panels in Figure \ref{fig:3}). 
The first supernova explosion also occurs at this time and thus 
the number of ionizing photons emerging from the central cluster begins to 
decay rapidly afterwards. Model LDSc presents the shell structure at a later 
time, $t = 5$~Myr, when all ionizing photons are absorbed in the inner 
segments of the shell and thus the outer skin of the shell remains neutral. 

The conditions for model LDL assume a leaky bubble model 
\citep[see, for example,][]{Matzner2002, Harper-ClarkMurray2009}. In this case, the thermal
pressure inside the wind-driven bubble drops below the \citet{Weaveretal1977} model predictions 
due to the escape of hot shocked-wind plasma through holes in the wind-driven 
shell. In this case individual bow shocks around the shell fragments should 
merge to create a coherent reverse shock near the contact discontinuity, or
inner side of the broken shell \citep[see][]{TenorioTagleetal2006,RogersPittard2013}. We thus assume that the minimum driving force on 
the shell in the leaky bubble model is determined by the cluster wind ram 
pressure at the shell location and can never fall below such value
(\citetalias{SilichTenorioTagle2013}). Hereafter we will assume 
that in the leaky case the transition from energy to momentum dominated 
regimes occurs at $0.13$~Myr, just after the shell cools down and begins to 
absorb ionizing photons. Equations (\ref{eq2a}-\ref{eq2c}) are replaced with 
equations (\ref{eq13a} - \ref{eq13c}) at this time. Certainly, this time is 
arbitrary, but warrants the maximum possible effect of radiation pressure.

The density and thermal pressure distributions within and at both sides of the
shell in the leaky case are shown on the right-hand panels of Figure 
\ref{sec:3}. Here
the top middle and bottom panels correspond to models LDLa, LDLb and LDLc and thus
present the density, thermal pressure and ram pressure profiles at the same 
evolutionary times $t = 1$, $3.3$ and $5$ Myr, respectively. The size of the leaky
shell is smaller and its thickness larger than those predicted by the standard
bubble model (model LDS) and the difference grows with time (compare the right 
and left-hand panels in Figure \ref{fig:3}).  Note also that the leaky 
shell is not able to trap all ionizing photons and form an outer neutral 
skin for a much longer time (in this case $\tau_{trap} \approx 5$~Myr). 
This is because in the leaky bubble model the driving pressure and 
thus the ionized gas density at the inner edge of the shell are much smaller 
than those in the standard case (LDS). 

The expectations resulting from calculations of the ionized gas 
distribution in static configurations with low pressure central cavities
(section \ref{sec:2}) had been that radiation pressure would lead to a non 
homogeneous thermal pressure and  density distributions inside the 
wind-driven shell. Both, density and thermal pressure should grow from a low 
value at the inner edge of the shell to a maximum value at the outer edge, 
as in the high density static models (see section \ref{sec:2}). The 
calculations however, do not show such large enhancements in density and in 
the leading shock driving pressure relative to that at the inner edge of the 
shell. The density enhancement  is about $\sim 1.04$ and $\sim 1.09$ 
at $1$~Myr, $\sim 1.07$ and $\sim 1.18$ at $3.3$~Myr and $\sim 1.04$ 
and $\sim 1.12$ at $5$~Myr in the standard and the leaky bubble model, 
respectively (see the left-hand and right-hand panels of Figure \ref{fig:3}). 
We then provided similar calculations for an order of magnitude less 
massive cluster (10$^5$ M$_{\sun}$) and did not find significant difference 
with the above results. In all calculations with a 10$^5$ M$_{\sun}$ cluster 
the density enhancement does not exceed $\sim 1.1$ despite radii of the shells
differ significantly from those obtained in the more energetic models LDS and 
LDL. These results demonstrate how significantly the inner boundary condition
(the value of thermal pressure at the inner edge of the HII region) may
change the ionized gas density distribution. They also imply that the impact 
from radiation pressure on the dynamics of shells formed by massive young 
stellar clusters embedded into a low density ambient medium is not significant 
throughout their evolution even if all of the hot plasma leaks out from the 
bubble interior into the surrounding medium. Consequently, allow for the use 
of equations (\ref{eq2a}-\ref{eq2b}) and (\ref{eq13a}-\ref{eq13b}), ignoring 
the impact of the starlight momentum. 

\subsection{Shells evolving in a high density ISM}

\begin{figure*}[htp]
\plotone{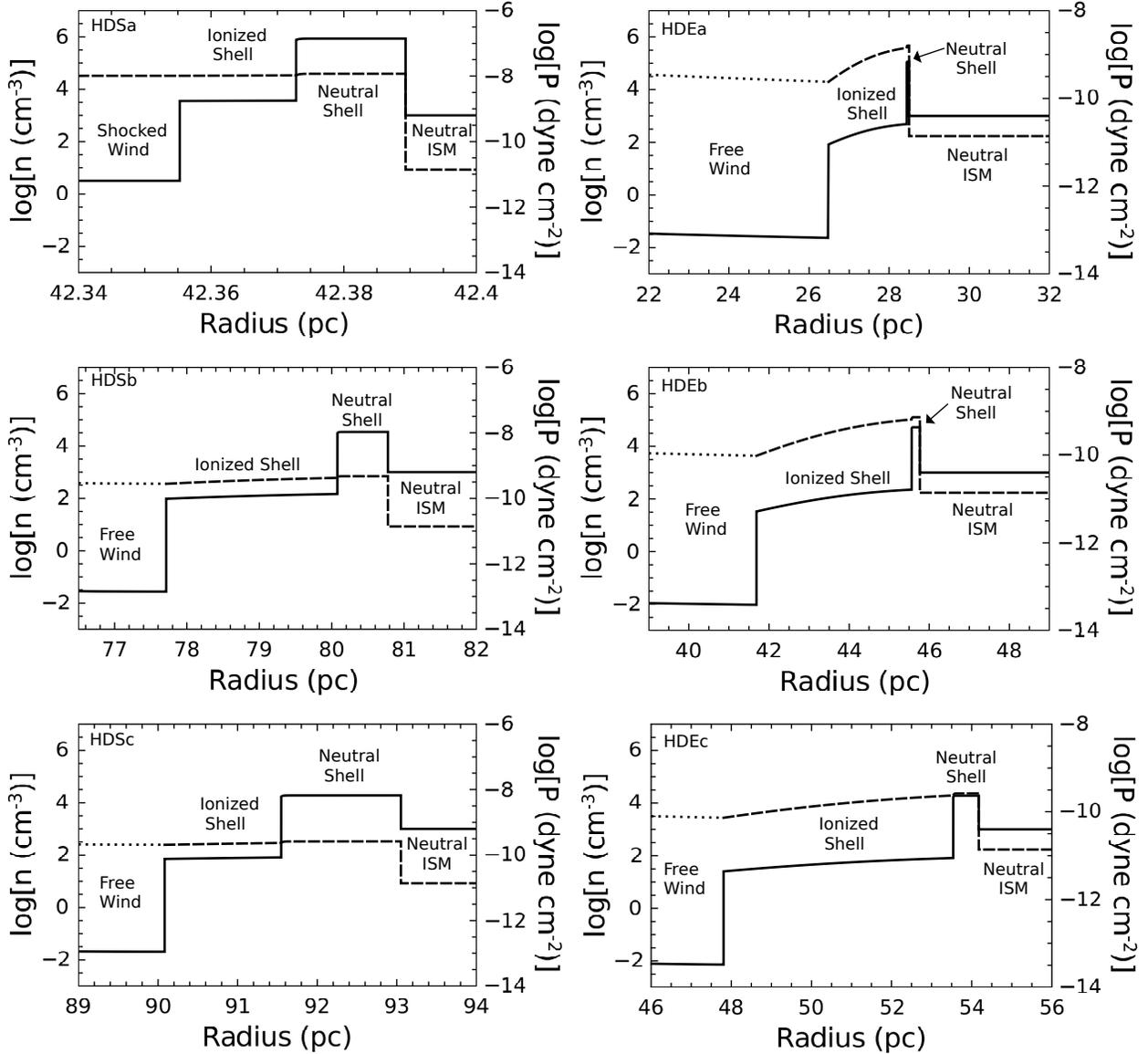}
\caption{The wind-blown shell structure for a high-density 
environment. The left-hand column shows the results for models HDSa (top panel),
HDSb (middle panel) and HDSc (bottom panel). The right-hand column displays the 
results for models with a low cluster heating efficiency: HDEa (top panel), 
HDEb (middle 
panel) and HDEc (bottom panel). Solid lines correspond to the radial density 
distribution (left axis) for the free wind, shocked wind, ionized shell, 
neutral shell and the ambient ISM. Long-dashed and short-dashed lines depict 
the radial thermal and ram pressure distributions (right axis), respectively.
The Mach number for the HDS models a, b and c is 23.9,
8.2 and 5.8, respectively, while for the HDE models a, b and c is 12.5, 
3.9 and 3.0.}
	\label{fig:4}
\end{figure*}

The high-density models (Table \ref{tab:2}, models HDS and HDE) are evaluated at the same 
dynamical times: $t $= 1, 3.3 and 5~Myr and are displayed in Figure 
\ref{fig:4}. In these cases the model predicts that the transition from energy 
to the momentum dominated regime occurs at much earlier times (see equation 
\ref{eq15}). For example, in the case of model HDS, $\tau_{tran} \approx 
1.58$~Myr. Thus, models HDSb and HDSc correspond to a shell expanding in the 
momentum dominated regime. The size of the shell in this case is much smaller 
than when it expands into a low density ISM, however the shell is much denser 
and thus recombines faster. Therefore in the high density cases the 
ionizing radiation is not able to photoionize the whole shell from the very 
early stages of the bubble evolution (see the top left-hand panel 
in Figure \ref{fig:4}). The density in the ionized shell drops when the 
transition to the momentum-dominated regime occurs. This allows the central 
cluster to photoionized a larger fraction of the swept-up material. Therefore 
the relative thickness of the ionized shell increases between the 1~Myr and 
3.3~Myr (compare panels HDSa and HDSb in Figure \ref{fig:4}). After 3.3~Myr the 
number of ionizing photons decreases rapidly (Figure \ref{fig:2}) and the 
relative thickness of the ionized shell becomes smaller again despite the 
drop in driving pressure and the consequent drop in the ionized gas density 
(see panel HDSc). The density gradient also reaches the maximum value at 
$3.3$~Myr and then drops at latter times.
The density (and thermal pressure) gradient across the ionized shell in the
high density models is larger than in the low density cases. For example, the 
enhancement of density relative to that at the inner edge of the shell in 
model HDSa is $\sim 1.14$, in model HDSb is $\sim 1.67$ and in model HDSc $\sim 
1.25$ (see the left-hand panels in Figure \ref{fig:4}). This is because the 
inner radius of the ionized shell in the high density case is smaller and 
thus the impact that radiation pressure provides on the shell is larger.

The right-hand panels in Figure \ref{fig:4} present the results of the 
calculations when the driving cluster has a low heating efficiency (models 
HDEa, HDEb and HDEc). These calculations were motivated by the discrepancy 
between the \citet{Weaveretal1977} model predictions and the observed sizes 
and expansion velocities of the wind-blown bubbles known as ``the growth-rate 
discrepancy'' \citet{Oey1996} or ``the missing wind problem'' 
(\citealt{Freyeretal2006, Dopitaetal2005,Smithetal2006,Silichetal2007,
Silichetal2009}) and by the fact that at the initial stages of the bubble 
evolution the star cluster mechanical luminosity still does not reach the 
average value adopted in our calculations (see Figure \ref{fig:2}). The 
heating efficiency may also be small if the kinetic energy of stellar winds 
is converted to turbulence and radiated away in young stellar clusters 
\citep{Bruhweileretal2010}. At later stages of evolution
a low heating efficiency may be physically justified by assuming mass 
loading of the matter left over from star formation, as in 
\citet{Wunschetal2011}, or an oversolar metallicity of the SN ejecta what 
enhances the cooling rate, as in \citet{TenorioTagleetal2005}.
More recently, a low heating efficiency has been shown to also arise from the 
consideration of a continuous presence of dust within the cluster volume, 
dust produced within the ejecta of the multiple core-collapsed SN 
expected in young clusters \citep[see][]{TenorioTagleetal2013}.
In this case, 
we keep the values of $L_i$, $L_n$ and $Q_0$ equal to those predicted by the 
Starburst99 synthetic model for a $10^6$\Msol \, cluster, but instead of 
using $L_{mech}=10^{40}$~erg s$^{-1}$, as in our models HDSa - HDSc, we use an 
order of magnitude smaller mechanical luminosity: $L_{mech} = 
10^{39}$~erg s$^{-1}$. The transition to the momentum dominated regime in
this case occurs at $\approx 0.84$~Myr. The relative thickness of the ionized 
shell is much larger than that in model HDS as the size of the shell is 
about two times smaller and thus the flux of the ionizing radiation is about 
four times larger than in model HDS (compare the left-hand and right-hand panels
in Figure \ref{fig:4}). 
This leads to the largest calculated enhancement in the shell density (and 
thus thermal pressure) relative to that at the inner edge of the shell 
which is: $\sim 6.41$ at $t=1$~Myr, $\sim 7.26$ at $3.3$~Myr and $\sim 
3.47$ at $5$~Myr. These results imply that radiation pressure must be 
taken into consideration in calculations with low heating efficiency and
that \citet{Weaveretal1977} model (equations \ref{eq2a}-\ref{eq2b} and 
\ref{eq13a}-\ref{eq13b}) must be corrected in this case. The radiation 
pressure may also contribute to the shell dynamics at very early stages 
(before 3~Myr) of the wind-blown bubbles evolution (see also Figure 3 in 
\citetalias{SilichTenorioTagle2013}). Similar results were obtained for the 
less massive ($10^5\Msol$) clusters. In this case the maximum enhancement of 
density is $\sim 1.43$ in the standard (HDS) case and $\sim 4.68$ in the low 
heating efficiency (HDE) model, respectively.

\begin{figure*}[ht]
\plotone{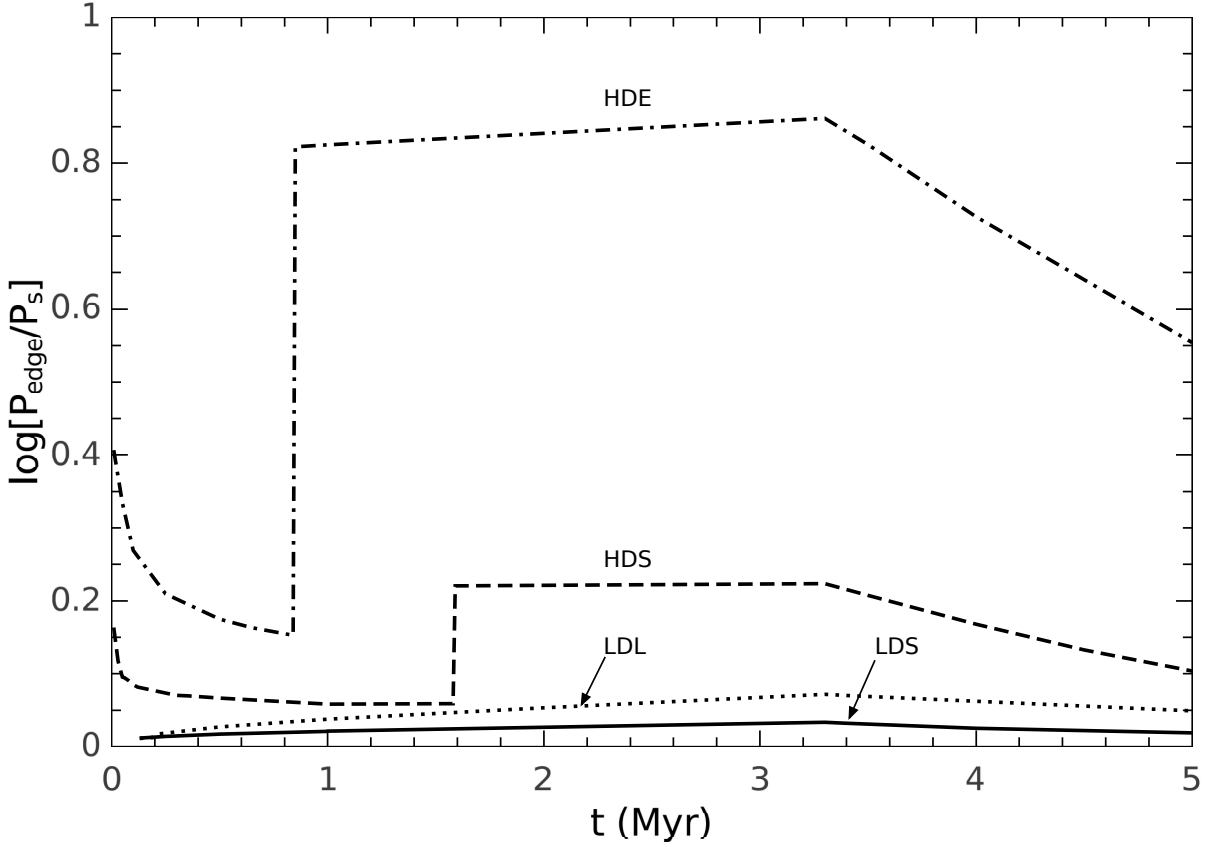}
\caption{The $P_{edge}/P_s$ ratio time evolution. The solid, dotted, dashed
and dash-dotted lines display the logarithm of the $P_{edge}$ over $P_s$ ratio,
$\log{P_{edge}/P_s}$, at different times $t$ in the case of models LDS, LDL, HDS and 
HDE, respectively.}
	\label{fig:5}
\end{figure*}

The time evolution  of the thermal pressure excess, $P_{edge}/P_s$, where 
$P_{edge}$ and $P_s$ are the values of the thermal pressure behind the leading
shock and at the inner edges of the wind-driven shell, is shown in Figure 
\ref{fig:5}. In the high density models (dashed and dash-dotted lines) this 
ratio decreases first as the flux of ionizing 
energy at the inner edge of the shell drops faster (as $R_s^{-2}$) than 
thermal pressure in the shocked wind region which drops as $R_s^{-4/3}$ (see 
equation \ref{eq2b}). It then grows to a larger value when the hydrodynamic 
regime changes from the energy to a momentum-dominated expansion and the wind 
pressure at the inner edge of the shell drops abruptly. After this time both, 
the flux of radiation energy and the wind ram pressure at the inner edge 
of the shell drop as $R_s^{-2}$. The $P_{edge}/P_s$ ratio then grows slowly as 
the number of non-ionizing photons absorbed by the outer neutral shell 
increases with time. The slow increase of the $P_{edge}/P_s$ ratio continues
until the number of ionizing photons begins to drop  after the first supernova 
explosion at 3.3~Myr when the $P_{edge}$ over $P_s$ ratio reaches 1.67 
($\log{P_{edge} / P_s} \approx 0.22$) in the case of model HDSb and 7.26 
($\log{P_{edge} / P_s} \approx 0.86$) in the case of model HDEb. 

In the low density cases (solid and dotted lines) the swept-up shell is not 
able to absorb all ionizing photons until it grows thick enough and therefore 
the number of ionizing photons trapped inside the completely ionized shell 
grows continuously until the first supernova explosion at 3.3~Myr. This 
compensates the $R_s^{-2}$ drop of the ionizing energy flux and leads to a 
continuously growing $P_{edge}/P_s$ ratio at this stage. However, in the 
standard (solid line) case and leaky (dotted line) bubble model this ratio 
remains always smaller than $\sim 1.7$. In the low density models LDS and LDL 
it is even smaller (less than 1.2) and is below the upper limit obtained
in \citetalias{SilichTenorioTagle2013}. This is because in the low density 
cases wind-driven shells absorb only a fraction of the star cluster bolometric 
luminosity. 

\begin{figure*}[htp]
\plotone{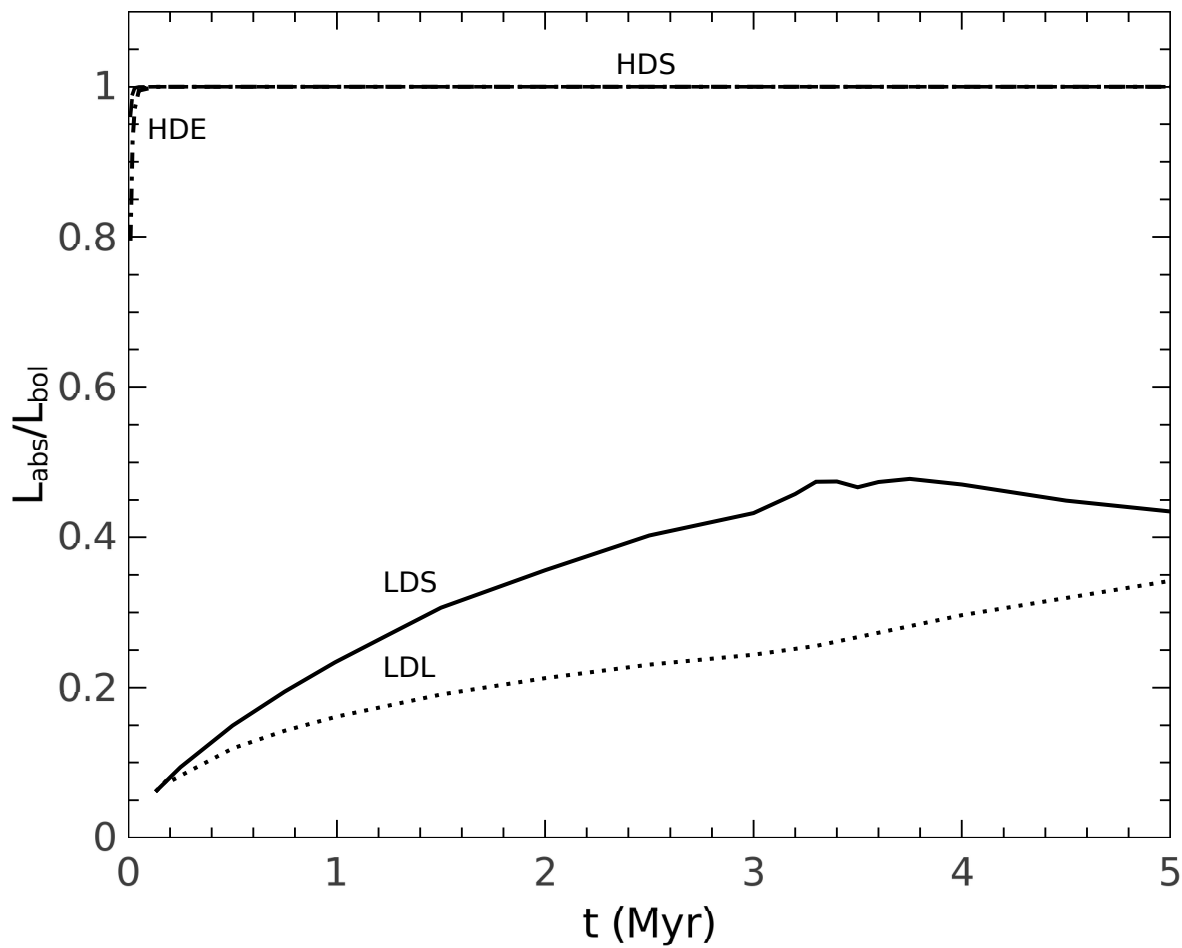}
\caption{Fraction of the star cluster bolometric luminosity trapped
within the shell as a function of time. The solid, dotted, dashed and 
dash-dotted lines display the $L_{abs}$ over $L_{bol}$ ratio at different 
evolutionary times $t$ for models LDS, LDL, HDS and HDE, respectively. Note that
dashed and dash-dotted lines overlap into a single horizontal line
$L_{abs} / L_{bol} = 1$ at the earliest stages of the shell evolution.}
        \label{fig:6}
\end{figure*}

\begin{figure*}[htp]
\plotone{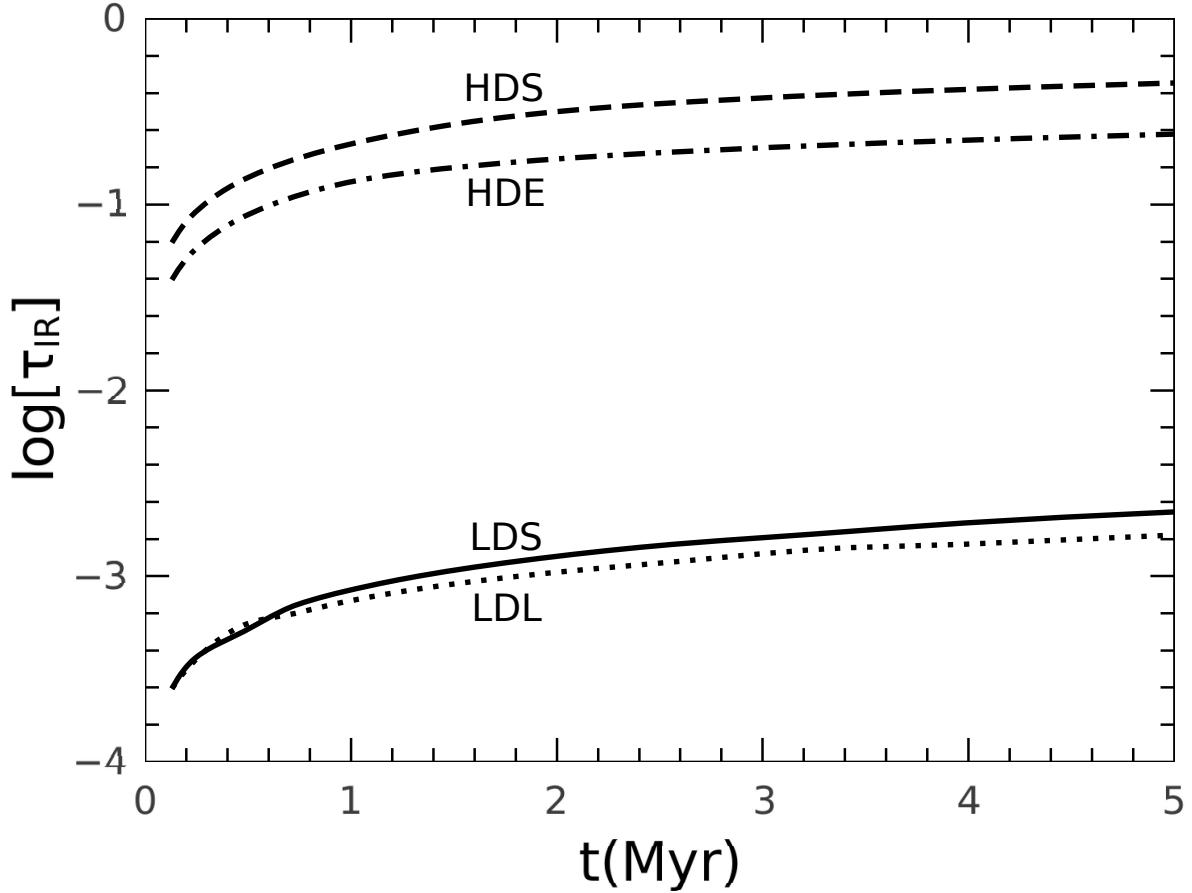}
\caption{The star cluster wind-driven shell optical depth for the IR radiation
as a function of time. The solid, dotted, dashed and dash-dotted lines show
$\tau_d$ for models LDS, LDL, HDS and HDE, respectively.}
        \label{fig:7}
\end{figure*}

The fraction of the star cluster bolometric luminosity trapped 
within a shell as a function of time in models LDS, LDL, HDS and HDE
is shown in Figure \ref{fig:6} by solid, dotted, dashed and dash-dotted lines,
respectively. Note that dashed and dash-dotted lines overlap rapidly as in 
the high density calculations expanding shells absorb all available (ionizing 
and non ionizing) photons at the very beginning of their time evolution
(at $t << 1$~Myr). However, even in this case the shell remains optically
thin to the IR photons re-emitted by dust grains, as shown in Figure 
\ref{fig:7}. Here we adopted for the dust opacity $\kappa_d = 2.3$~cm$^2$ 
g$^{-1}$ \citep[see Table 1 in][]{Novaketal2012} and calculated the optical 
depth for the IR radiation as $\tau_{IR} = \kappa_d \Sigma_s$, where the 
column density of the shell, $\Sigma_s$, is $\Sigma_s = \rho_{ISM} R_s / 3$. 
The amplification of radiation pressure by the multiple re-emitted IR photons 
which is $\sim \tau_{IR} L_{bol} / c$ 
\citep[see][]{Hopkinsetal2011,KrumholzThomson2012} thus remains less than 
unity. In all our calculations, the amplification factor never exceeds 2, 
even if one uses a larger dust opacity, $\kappa_d = 5$~cm$^2$  g$^{-1}$ 
adopted by \citet{Hopkinsetal2011}. This implies that the star cluster 
wind-driven shells expand in the radiation momentum rather than in the 
radiation energy dominated regime \citep[see][for the detailed discussion of 
the two limiting cases]{Falletal2010,KrumholzThomson2012}.

We also computed how radiation pressure affects the density and thermal 
pressure distribution in the case when the exciting cluster is embedded
into a low ($n_{ISM} = 1$~cm$^{-3}$) density ISM and has a low heating 
efficiency and in the case of a leaky shell moving into a high 
($n_{ISM} = 1000$~cm$^{-3}$) density medium. We found a little difference
between these calculations and models LDL and HDS, respectively. For example, 
the enhancement of density from the inner to the outer edge of the shell
in the low density calculations with a 10\% heating efficiency is about 1.13,
1.2 and 1.11 at 1~Myr, 3.3~Myr and 5~Myr, whereas in the leaky bubble model 
LDL it is $\sim 1.1$, $\sim 1.19$ and $\sim 1.12$, respectively. In the case 
when a leaky shell expands into a high density medium, the enhancement of 
density is : $\sim 1.61$ at 1~Myr, $\sim 1.67$ at 3.3~Myr and $\sim 1.25$ at 
5~Myr, whereas in model HDS it is $\sim 1.14$, $\sim 1.67$ and $\sim 1.25$, 
respectively, and thus the only difference between the last two models is 
that the transition from energy to momentum dominated regimes occurs at 
different times. Therefore we do not present the detailed description of these 
calculations in our further discussion.

\subsection{Comparison to other models and observations}

Having the exciting cluster parameters and the distribution of the ionized
gas density in the surrounding shell, one can obtain the model predicted 
values for diagnostic parameters often used in observations and compare them 
to the typically observed ones. In this section we first calculate the values 
of the ionization parameter and then put our results onto a diagnostic diagram
proposed by \citet{YehMatzner2012} which allows one to conclude if radiation 
or the wind dynamical pressure dominates the dynamics of the ionized gas 
around young stellar clusters.

\begin{figure}[htp]
\plotone{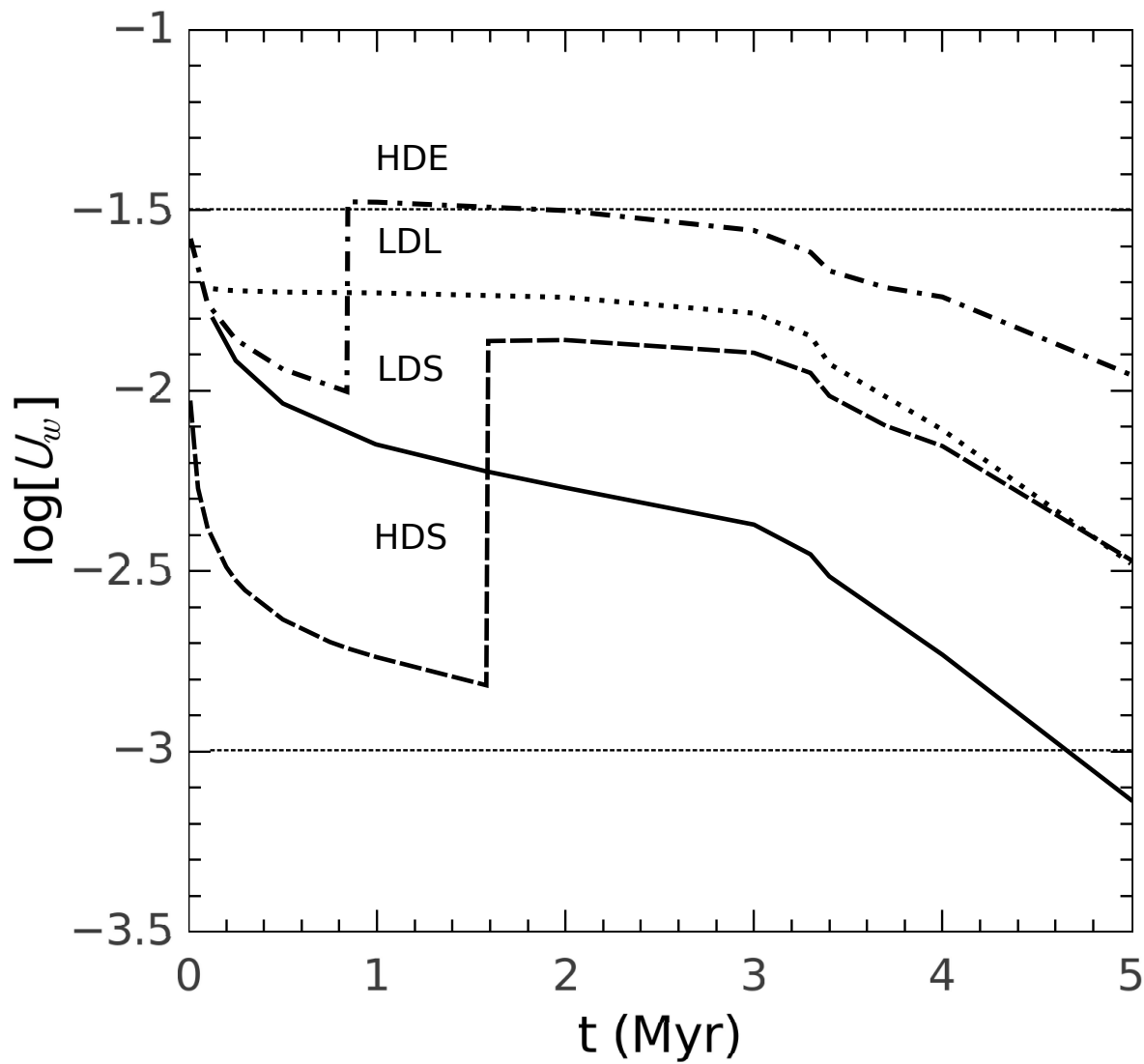}
\caption{The ionization parameter time evolution. The solid, dotted, dashed 
and dash-dotted lines correspond to models LDS, LDL, HDS and HDE, respectively 
(see Table \ref{tab:2}).
The horizontal lines display the range of typical values for the ionization
parameter found in local starburst galaxies \citep[see][]{RigbyRieke2004}.}
	\label{fig:8}
\end{figure}

The ionization parameter $U$ is defined as the flux of ionizing photons per 
hydrogen atom. It is directly related to the state of ionization and to the 
radiation pressure over gas thermal pressure ratio and is usually 
calculated at the inner edge of the ionized medium 
\citep[e.g.][]{Dopitaetal2005}:
\begin{eqnarray}
\label{eq:U1}
U=\frac{Q_{0}}{4\pi n R_{s}^2 c} = \frac{\mu_i}{\mu_a}\frac{k T_{i}}{\langle h\nu\rangle_i}\frac{P_{rad}}{P_{HII}} .
\end{eqnarray}
The ionization parameter may be measured observationally from the emission 
line ratios \citep[e.g.][and references therein]{RigbyRieke2004,
Snijdersetal2007,YehMatzner2012} and thus is a powerful tool to measure the 
relative significance of the radiation and gas thermal pressure around young 
stellar clusters. However, the number of ionizing photons varies radially within HII 
regions and therefore the measured values of $U$ are weighted by the density 
distribution in the ionized nebula. This led \citet{YehMatzner2012} to propose
as a relevant model parameter
\begin{equation}
\label{eq:U2}
U_w = \displaystyle \frac{\int 4 \pi r^2 n^2 U(r) \mbox{d}r }{\int 4 \pi 
        r^2 n^2 \mbox{d}r} ,
\end{equation}
where the integrals are evaluated from the inner to the outer edge of the HII
region. In our approach, we have neglected the presence of any neutral 
gas and dust able to deplete the radiation field in the free and hot shocked 
wind regions and thus assumed that all the photons produced by the 
star cluster are able to impact the shell. The integrals in equation 
\ref{eq:U2} thus were evaluated with the lower and upper limits 
$R_{s}$ and $R_{HII}$, respectively. Here we make use of our models
to obtain the ionized gas density distribution within wind-driven shells 
expanding into different interstellar media and calculate the ionization 
parameter $U_w$ at different times $t$. The results of the calculations are 
presented in Figure \ref{fig:8}. One can note, that
the time evolution of the ionization parameter $U_w$ in the wind-driven bubble
model is complicated as it depends not only on the varying incident radiation,
but also on the hydrodynamics of the wind-driven shell. In all cases the 
value of $U_w$ drops first as the wind-driven shell expands and the photon 
flux at the inner edge of the shell drops accordingly. In the standard case 
(LDS, solid line) the value of $U_w$ drops continuously but turns to decrease 
faster after the first supernova explosion as since that time the flux of
incident photons per unit area drops not only because of the shell expansion, 
but also because of the reduced value of $Q_0$. In the high density model HDS 
(dashed line) the value of the ionization parameter increases by about an 
order of magnitude after the transition to the momentum-dominated regime as 
when the transition occurs, the wind pressure and the ionized gas density
at the inner edge of the shell drop, what enhances the value of $U_w$ 
significantly (see equations \ref{eq:U1} and \ref{eq:U2}).
The value of the ionization parameter 
then remains almost constant until the first supernova explodes at about 
3.3~Myr as at this stage both, the flux of ionizing photons and the ram 
pressure of the wind at the inner edge of the shell drop as $R_s^{-2}$ and 
thus the radiation over the dynamical pressure ratio depends only on the 
$L_{bol} / L_{mech}$ ratio \citepalias[see][]{SilichTenorioTagle2013} which in our 
calculations does not change much at this stage. After 3.3~Myr the value of 
the ionization parameter drops as the number of massive stars and the number 
of available ionizing photons $Q_0$ decline rapidly. The behavior of $U_w$ 
in the leaky (model LDL, dotted line) and low heating efficiency 
(model HDE, dash-dotted line) cases is very similar to that in the high density 
case HDS. The only difference is that the transition to the momentum-dominated 
regime in these cases occurs at earlier times and the maximum values of the 
ionization parameter are larger than that in model HDS. One can also note that 
the ionization parameter reaches the maximum possible value, 
$log U_w \approx - 1.5$, in the low heating efficiency model HDL and that the
model predicted values of the ionization parameter fall into the range of
typical values found in local starburst galaxies: $-3 \le log U_w \le -1.5$, 
\citep[see Figure 10 in][]{RigbyRieke2004}. The larger values of the ionization
parameter \citep[e.g.][]{Snijdersetal2007} either require a lower heating 
efficiency, as was also claimed in \citet{Dopitaetal2005}, or a more 
complicated physical model than a single ionized shell formed by a young 
stellar cluster \citep[see the discussion in][]{Snijdersetal2007}.

\begin{figure}[htp]
\plotone{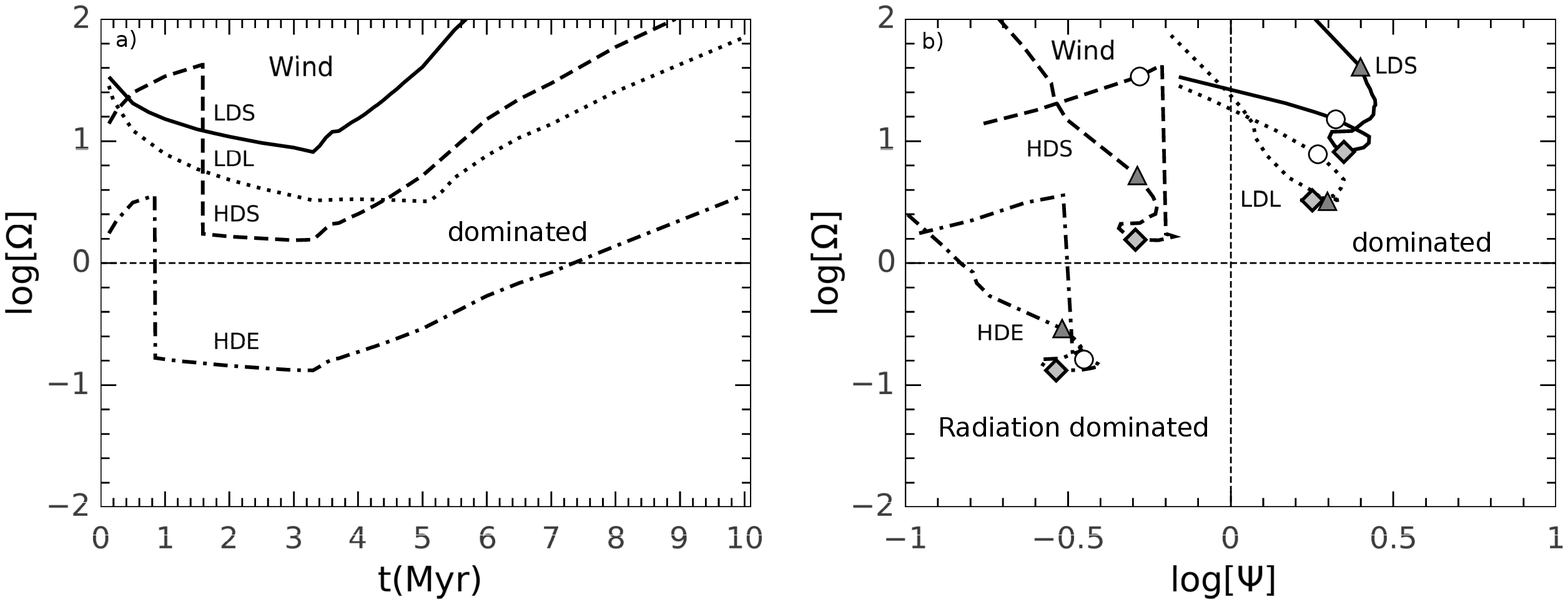}
\caption{Evolutionary tracks of the expanding ionized shells in the diagnostic
parameter space. The evolution of the diagnostic parameter $\Omega$ (see text)
is presented in panel a. Panel b displays the location of the expanding shell
exposed to the radiation from the central cluster in the $\log{\Omega} - 
\log{\Psi}$ diagram at different times $t$. The solid, dotted, dashed and 
dash-dotted lines correspond to models LDS, LDL, HDS and HDE, respectively. 
The circles, diamonds and triangles mark the evolutionary times of 
1 Myr, 3.3 Myr and 5 Myr, respectively.} 
	\label{fig:9}
\end{figure}

Finally, we put our results onto a diagnostic diagram proposed by 
\citet{YehMatzner2012} in order to show where physically motivated models
are located in this diagram. For example, their model with more than an
order of magnitude increasing density (see Figure 7 in their paper), 
$L_i = 10^{42}$~erg s$^{-1}$, $\log{\Phi} = -1.09$ and $\log{\Omega} = 
-1.56$ corresponds, according to our calculations, to a very compact 
($R_{HII}$ less than 3~pc) and very dense ($n_s$ is a few hundred particles 
per cm$^{3}$) shell at the age of 2~Myr what implies that the HII region 
is quasi-static and requires a very low star cluster heating efficiency and 
a large confining (thermal/turbulent) pressure in the ambient ISM 
\citep[see][]{Smithetal2006,Silichetal2007,Silichetal2009}. 
Two-dimensional parameter space introduced by \citet{YehMatzner2012} is
related to the compactness of the HII region (parameter $\Psi$) and to 
the relative strength of different driving forces (parameter $\Omega$).
Parameter $\Psi$ is defined as the $R_{HII} / R_{ch}$ ratio, where $R_{HII}$ 
is the radius of the ionization front (in our case this is the radius of the 
outer edge of the ionized shell) and $R_{ch}$ is the radius of a uniform 
density Str\"omgren sphere whose thermal pressure is equal to the maximum
possible unattenuated radiation pressure at the edge of the HII region 
$P_{rad} = L_{bol} / 4 \pi c R^2_{st}$:
\begin{equation}
\label{eq:rch}
R_{ch} = \frac{\beta_2 \mu^2_a L^2_{bol}}{12 \pi \mu^2_i (k T_i c)^2 Q_0} .
\end{equation}
Parameter $\Omega$ is related to the volume between the ionization front
and the inner edge of the HII region and to the values of thermal pressure 
at its inner and outer edges:
\begin{equation}
\label{eq:omega1}
\Omega = \frac{P_s R^3_s}{P_{edge} R^3_{edge} - P_s R^3_s} .
\end{equation}
We obtain parameter $\Omega$ by calculating the volume between the outer 
and the inner edge of the ionized shell and the values of thermal pressure 
$P_s$ and $P_{edge}$ even at earlier stages of models LDS and LDL when the 
ionized shell is still embedded into an extended diffuse HII region. As long 
as the ionized shell is thin, parameter $\Omega$ is:
\begin{equation}
\label{eq:omega2}
\Omega \approx 4 \pi c R^2_s P_s / L_{bol} ,
\end{equation}
and thus measures the wind dynamical over the radiation pressure ratio
(the shell moves in the radiation-dominated regime if $\log{\Omega} < 0$
and in the wind-dominated regime if $\log{\Omega} > 0$).
In all static models discussed in section \ref{sec:2} the parameter 
$\Omega$ is very small ($\log{\Omega}\sim-15$) what implies that
radiation pressure controls the ionized gas distribution in all
static configurations with low-pressure central cavities. In the 
wind-blown cases the parameter $\Psi$ is a function of time 
as both radii, $R_{HII}$ and $R_{ch}$, change with time. Therefore it is 
instructive to show first  how parameter $\Omega$ changes with time. This is 
shown in Figure \ref{fig:9} , panel a. Panel b in this figure displays the 
evolutionary tracks of our models in the $\Omega - \Psi$ parameter space. 
The initial points for models LDS, LDL, HDS and HDE were calculated at the 
star cluster age of 0.13~Myr. The initial values of the normalization radius 
$R_{ch}$ then are: $\sim 72$~pc in model LDL and $\sim 70$~pc in models LDS, 
HDS and HDE, respectively. 
As both star cluster parameters, $L_{bol}$ 
and $Q_0$, change with time, the value of $R_{ch}$ also changes with time
significantly and by 10~Myr reaches $\approx 720$~pc. In cases LDS and LDL 
parameter $\Omega$ grows continuously (see panel a, solid and dotted lines). 
In the high density cases parameter $\Omega$ drops drastically when the 
transition occurs to the momentum dominated regime, then slightly declines 
and increases again after the first supernova explosion as the number of the 
ionizing photons then drops rapidly.  
The strong time evolution of $R_{ch}$ leads to the intricate tracks of the
ionized shells in the $\log{\Omega} - \log{\Psi}$ diagram (see panel b). 
In the low density models LDS and LDL the tracks go to the left and up because
the normalization radius $R_{ch}$ grows with time faster than the radius of
the shell and thus the ionization front radius $R_{HII}$. In the high density 
cases HDS and HDE the tracks are more intricate. They first go to the right, 
then drop down when the transition to the momentum dominated regime occurs, 
make a loop and finally go back to the left and up. 

Thus, in the low density cases the impact of radiation pressure on the shell 
dynamics is always negligible and declines with time. In the high density 
model HDS the contribution of radiation pressure to the shell dynamics becomes 
more significant when the shell makes a transition from the energy to the
momentum dominated regime. However, in this case parameter $\log{\Omega}$ 
also remains positive and thus in all models with a 100\% heating efficiency 
the shells expand in the wind-dominated regime. Parameter $\log{\Omega}$ 
falls below a zero value only in the low heating efficiency case HDE. Thus,
only in this case radiation pressure may dominate the shell dynamics.
The radiation dominated phase lasts from the beginning of the momentum 
dominated regime at $\sim 0.85$~Myr till $\sim 7.36$~Myr (see panel a).
This implies that radiation pressure may dominate the dynamics of the gas
around young stellar clusters either at early stages of evolution 
(before $\sim 3$~Myr) or if the major fraction of the star cluster mechanical
luminosity is dissipated or radiated away within the star cluster volume and 
thus the energy of the star cluster driven winds is significantly smaller
than what star cluster synthetic models predict. However, even if this is the
case, radiation pressure will dominate only if the exciting cluster is 
embedded into a high density ambient medium.

\section{Summary}
\label{sec:5}

1. Radiation pressure may strongly affect the structure of static, dusty 
   HII regions. However, the impact that star cluster winds provide on the
   flows and the strong time evolution of the ionizing photon flux and the 
   star cluster bolometric luminosity lead to a more intricate picture. 

2. In a more realistic model, the impact of radiation pressure on the 
   expanding shell crucially depends on the strength of the star cluster 
   wind at the inner edge of the shell and thus on the hydrodynamic regime 
   of the shell expansion and on the star cluster age and heating 
   efficiency. 

3. Radiation pressure may affect the inner structure and the dynamics
   of the wind-driven shell only at the earliest stages of evolution 
   (before $\sim 3$~Myr, when the $L_{bol}$ over $L_{mech}$ ratio is still
   larger than that used in our calculations), or if a major fraction of 
   the star cluster mechanical luminosity is dissipated or radiated away 
   within the star cluster volume and thus the star cluster mechanical 
   energy output is much smaller than star cluster synthetic models predict. 
   However, even in these cases radiation effects may be significant only if 
   the exciting cluster is embedded into a high density ambient medium.
   
4. The impact that radiation pressure provides on the dynamics and inner
   structure of the wind-driven shell is always negligible during the advanced 
   stages of evolution as the radiation energy flux declines rapidly after 
   the first supernovae explosion whereas the mechanical power of the cluster 
   does not. 

5. The calculated values of the density weighted ionization parameter $U_w$ 
   fall into the range of typical values found in nearby starburst 
   galaxies ($-3 \le$ log$U_w \le -1.5$). The larger values of the ionization
   parameter sometimes detected around very young stellar clusters require 
   either a lower heating efficiency, or a more complicated than a single 
   ionized shell physical model.

6. The model location in the $\log{\Omega} - \log{\Psi}$ diagnostic diagram 
   proposed by \citet{YehMatzner2012} strongly depends on the evolutionary 
   time $t$ what leads to intricate evolutionary track patterns. The standard 
   wind-driven and leaky bubble model are located in the upper segments in 
   this diagram where HII regions evolving in the thermal pressure dominated 
   regime settle in. The only model whose evolutionary track passes through 
   the lower left corner where radiation pressure dominated HII regions are 
   located, is and only temporarily, that with a low heating efficiency. 

\acknowledgments

We thank our anonymous referee for a detailed report full of valuable 
comments and helpful suggestions which greatly improved the paper. SS also 
thanks M. Fall for a useful discussion of the radiation pressure 
amplification provided by the re-radiated IR photons. This study 
was supported by CONACYT, M\'exico through research grants 167169 and 131913. 

\bibliographystyle{apj2}
\bibliography{IonizedShells}

\end{document}